\documentclass[aps,pra,twocolumn,showpacs,amsmath,amssymb,floatfix]{revtex4}

\usepackage{graphicx} 
\usepackage{dcolumn} 
\usepackage{bm} 
\usepackage{epsfig}
\usepackage{amsmath}
\usepackage{amssymb}
\usepackage{float}
\usepackage{url}
\usepackage{feynmf}
\usepackage{color}
\usepackage{ulem}
\begin{document}
\newcommand{\bsy}[1]{\mbox{${\boldsymbol #1}$}} 
\newcommand{\bvecsy}[1]{\mbox{$\vec{\boldsymbol #1}$}} 
\newcommand{\bvec}[1]{\mbox{$\vec{\mathbf #1}$}} 
\newcommand{\btensorsy}[1]{\mbox{$\tensor{\boldsymbol #1}$}} 
\newcommand{\btensor}[1]{\mbox{$\tensor{\mathbf #1}$}} 
\newcommand{\tensorId}{\mbox{$\tensor{\mathbb{\mathbf I}}$}} 
\newcommand{\be}{\begin{equation}}
\newcommand{\ee}{\end{equation}}
\newcommand{\bea}{\begin{eqnarray}}
\newcommand{\eea}{\end{eqnarray}}
\newcommand{\e}{\mathrm{e}}
\newcommand{\arccot}{\mathrm{arccot}}
\newcommand{\arctanh}{\mathrm{arctanh}}

\title{Electromagnetic quantum shifts in relativistic Bose-Einstein condensation}
 
\author{D. M. Reis$^{1}$ and C. A. A. de Carvalho$^{1,2}$}

\affiliation{
$^1$Centro Brasileiro de Pesquisas F\'{\i}sicas - CBPF, Rua Dr. Xavier Sigaud 150, Rio de Janeiro - RJ, 22290-180, Brazil\\
$^2$Instituto de F\'{\i}sica, Universidade Federal do Rio de Janeiro - UFRJ, Caixa Postal  68528, Rio de Janeiro - RJ, 21945-972, Brazil}

\begin{abstract}
We compute deviations from ideal gas behavior of the pressure, density, and Bose-Einstein condensation temperature of a relativistic gas of charged scalar bosons caused by the current-current interaction induced by electromagnetic quantum fluctuations treated via scalar quantum electrodynamics. We obtain expressions for those quantities in the ultra-relativistic and nonrelativistic limits, and present numerical results for the relativistic case.

\end{abstract}

\pacs{71.10.Ca; 71.45.Gm; 78.20.Ci.}

\date{\today}

\maketitle

In a {\sl nonrelativistic} ideal gas of particles of mass $m$ obeying Bose-Einstein statistics, the energy of a particle is simply $E(\vec{p}\,)= \vec{p}\,^2/2m$. One may fix the number of particles of the gas, a conserved quantity, and compute it by integrating over the occupation numbers of excited states to extract the temperature where a macroscopic fraction of the particles starts condensing in the $\vec{p}=\vec{0}$ ground state, which does not contribute to the sum \cite{Landau-Lifchitz}. This Bose-Einstein condensation phenomenon has been observed in several physical systems, most recently in ultra-cold atoms \cite{Exp1, Exp2}. 

In a {\sl relativistic} ideal gas, the number of particles is no longer conserved. Instead, it is the number of particles $N_+$ minus the number of antiparticles $N_-$ that is conserved, an indication that antiparticles must be included in a relativistic treatment \cite{Haber-Weldon}. In fact, a comparison between the free energy of the gas without antiparticles \cite{CAC-Rosa} and that with particles and antiparticles shows that the latter is smaller, thus favored \cite{lieb}.  

In this letter, we study the charged relativistic Bose gas, which is made up of charged scalar bosons and antibosons of rest mass $m$ and energy $E_{\pm}(\vec{p}\,)= \pm \sqrt{\vec{p}\,^2 c^2+ m^2c^4}$ ($+$ bosons, $-$ antibosons) in a neutralizing background. A chemical potential $\mu \in [- m c^2, +m c^2]$ is used to fix the conserved charge, proportional to the number of bosons minus antibosons. This ideal gas also undergoes a phase transition, forming a Bose-Einstein condensate below a critical temperature $T_c$ \cite{Haber-Weldon, CAC-Rosa,lieb}.

Because of its {\sl electromagnetic} (EM) charge $e$, the gas will inevitably couple to EM quantum fluctuations that induce a current-current interaction. Thus, it may no longer be treated as ideal - one has to  treat it via scalar quantum electrodynamics at finite temperature and charge density, the microscopic theory that naturally incorporates such fluctuations. 

Normally, BEC is considered only for free boson gases. The inclusion of interactions, going now from an ideal gas into a real gas, in general is highly non trivial and several procedures have been developed. The influence of an interacting potential on the BEC critical temperature in the non-relativistic case has been discussed by several authors \cite{baym},\cite{pethick}. Here, however, interaction emerges from vacuum fluctuations in scalar electrodynamics. No interacting potential is introduced by hand. The idea is to integrate out the photon degrees of freedom, producing and effective interaction for the charged bosons.

Within that framework, we will show that the interaction changes the pressure, density, and critical temperature of condensation of the gas with respect to their ideal gas values by amounts that depend on the fine structure constant $\alpha = e^2/4\pi \hslash c$, and compute the deviations in the ultra-relativistic (UR), relativistic, and nonrelativistic (NR) regimes.  

Our calculation starts with the grand partition function of the system $\Xi={\rm Tr}\, e^{-\beta (\hat H - \mu \Delta \hat N)}$, $\Delta \hat N \equiv \hat N_+ - \hat N_-$, written as a functional integral over EM and scalar fields. Doing the quadratic integration over EM quantum fluctuations, we obtain an effective action for the scalars with an induced current-current interaction proportional to $\alpha$.

We treat the induced interaction as a perturbation in the remaining integral over the scalars, and compute a Feynman graph that gives the current-current expectation value. Finally, we integrate that expectation value, times the photon propagator, over photonic momenta.

From the thermodynamic potential $\Omega = - T \ln \Xi$, we obtain the pressure $P= -\Omega/V$, and the density $\eta= \Delta N/V = (\partial P/\partial \mu)_{T,V}$. Since, as before, the zero momentum state does not contribute to the sum, the condensation temperature comes from equating the integral over occupation numbers to $\eta$.  

In natural units \cite{units}, the action for Euclidean scalar quantum electrodynamics at finite temperature and density  is \cite{Kapusta}
\be
\label{action}
{S}= \int_{\Omega}\!\! d^4 x\,  \left [ \frac{1}{4}F_{\rho\sigma}F_{\rho\sigma}+\bar{D}_\rho\phi^* \bar{D}_\rho\phi+m^2\phi^*\phi \right ],
\ee
where $\int_{\Omega} d^4 x \equiv \int_0^\beta dx_4\int_V d^3 x$, $\bar{D}_\rho\phi=(\bar{\partial}_\rho-ie A_\rho)\phi$, and $\bar{\partial}_\rho\equiv(\partial_i, \partial_4-\mu)$. The grand partition function of the system may be expressed as a functional integral \cite{Kapusta}
\be
\Xi= \oint [d\phi^*] [d\phi] d\Sigma[A] \,  e^{-S[\phi^*, \phi, A]},
\ee
where $d\Sigma[A]$ is the gauge invariant measure for the EM field \cite{measure}. The integral symbol denotes a sum over field configurations of $A_\rho$, $\phi$, and $\phi^*$ whose value at $(\vec{x}, 0)$ is the same as at $(\vec{x}, \beta)$, boundary conditions that implement the trace.

We sum over the quantum fluctuations of the EM field, by doing the quadratic integral over $A_\rho$, and derive a current-current interaction. The grand partition function becomes
\be
\label{Zscalar}
\Xi=\oint [d\phi^*][d\phi] e^{-({S}_0 + {S}_I)},
\ee
where ${S}_0$ is the free bosonic action, obtained by setting $A_\rho = 0$ in Eq.(\ref{action}), and the interacting part is
\be
{S}_I [\phi] = - e^2\!\! \int_{\Omega}\!\! d^4 x \int_{\Omega}\!\! d^4 x' J_\rho(x) G_{\rho\sigma}^\phi (x - x') J_\sigma(x').
\ee
The current $J_\nu=i(\phi^*{\partial}_\nu\phi - \phi{\partial}_\nu\phi^* )$ interacts via the photon propagator in the background of the field $\phi$, whose inverse is $
(G_{\rho\sigma}^\phi)^{-1} \equiv \Gamma^\phi_{\rho\sigma}= \Gamma_{\rho\sigma} + 2e^2 |\phi|^2\delta_{\rho\sigma}$. 
For the free propagator, we have $\Gamma_{\rho\sigma}= -\partial^2\delta_{\rho\sigma}+(1- {\lambda}^{-1})\partial_\rho\partial_\sigma$, with $\lambda$ a gauge parameter. In momentum space,
\begin{equation}
\label{freeprop}
{G}_{\rho\sigma}(q)= \Gamma_{\rho\sigma}^{-1}(q) = \frac{\delta_{\rho\sigma}}{q^2}+\frac{(\lambda-1)}{q^2}\frac{q_\rho q_\sigma}{q^2}.
\end{equation}
Note that the term proportional to $e^2|\phi|^2$ is not present in $G_{\rho\sigma}(q)$. It comes from the coupling of two photons with two bosons in the interaction Lagrangian density $\mathcal{L}_\textrm{int}=e J_\mu A_\mu+e^2A^2|\phi|^2$. However, if we include the term $e^2 A^2 |\phi|^2$ through $G_{\rho\sigma}^\phi$, instead of using the free propagator, in the expression for the grand partition function $\Xi$ (see below), the order $e^2\sim \alpha$ extra term will contribute in order $e^4\sim \alpha^2$, which can be neglected, as we will show in the numerical results. 

We expand the grand partition function to first order to find
\be
\label{grand}
\Xi/\Xi_0 = 1 +\frac{e^2}{2}\!\! \int_\Omega d^4x \!\! \int_\Omega\!\! d^4y \,G_{\rho\sigma}(x-y) \langle J_\rho(x) J_\sigma(y) \rangle,
\ee
where $\Xi_0$ is the ideal gas grand partition function (${S}_I=0$), and the current-current expectation value is computed from the scalar theory given by Eq.(\ref{Zscalar})
\be
\label{correlation}
\langle J_\rho(x)J_\sigma(y)\rangle = \langle J_\rho(x)J_\sigma(y)\rangle_c+\langle J_\rho(x)\rangle \langle J_\sigma(y)\rangle, 
\ee
with $\langle J_\rho(x)\rangle = \delta_{\rho 4} \eta $. The second term in Eq.(\ref{correlation}) gives a $(\ln V/V)$ contribution to the pressure that vanishes in the thermodynamic limit. Besides, it does not contribute to the density either, since it is independent of $\mu$ \cite {finite size}. On the other hand, the connected part $\langle J_\rho(x)J_\sigma(y)\rangle_c \equiv {\cal J}_{\rho\sigma}(x-y)$ leads to
\be
\Xi/\Xi_0 = 1+ \frac{e^2V}{2}\!\!\!\!\sum_{n_q=-\infty}^{\infty}\!\!\int\!\!\frac{d^3q}{(2\pi)^3}{\cal J}_{\rho\sigma}(q_\mu) {G}_{\rho\sigma}(-q_\mu).
\ee
${\cal J}_{\rho\sigma}$  is given by the Feynman graph  
\begin{figure}[ht!]
	\begin{center}
		\includegraphics[width=25mm]{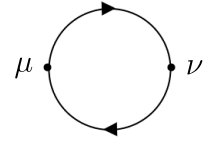}
\caption{\footnotesize Feynman diagram.}
		\label{fig:feynman}
	\end{center}
\end{figure}

\be
\label{sigma}
{\cal J}_{\rho\sigma} = T \!\!\sum_{n_p=-\infty}^{\infty}\!\!\int\!\! \frac{d^3p}{(2\pi)^3}\frac{(2\bar{p}_\rho+q_\rho)(2\bar{p}_\sigma+q_\sigma)}{(\bar{p}^2+m^2)[(\bar{p}+q)^2+m^2]},
\ee
with $q_4=2\pi n_q T$, $\bar{p}_4=2\pi n_p T+i\mu$. Due to current conservation, $q_\rho {\cal J}_{\rho\sigma}={\cal J}_{\rho\sigma} q_\sigma =0$, the gauge-dependent second  term of the propagator in Eq.(\ref{freeprop}) will not contribute. Introducing the trace ${\cal J}(q_\mu) \equiv {\cal J}_{\rho\rho} (q_\mu)$, we have
\begin{equation}
\label{Z}
\Xi/\Xi_0= 1 +\frac{e^2V}{2}\!\!\sum_{n_q=-\infty}^{\infty}\!\!\int\!\!\frac{d^ 3q}{(2\pi)^3}\frac{{\cal J}(q_\mu)}{q^2}.
\end{equation}

If we perform the Matsubara sum in Eq.(\ref{sigma}), and subtract the vacuum part ${\cal J}_{(0,0)}$, the medium contribution ${\cal J}_{m}={\cal J}_{(T,\mu)}-{\cal J}_{(0,0)}$ is given by
\be
\label{Jm}
{\cal J}_{m}(q_\mu)={\rm Re}\!\!\int\!\! \frac{d^3p}{(2\pi)^3}\frac{\bar{n}(p)}{\omega_p}\left[\frac{q^2+4pq -4m^2}{q^2+2pq}\right],\\
\ee
where $pq=i\omega_p q_4+\vec{p}\cdot\vec{q}$, $\omega_k=({\Vec{k}\,^2+m^2})^{1/2}$, and $\bar{n}(k)=n^+(k)+n^-(k)$,
\begin{equation}
    n^{\pm}(k)=\frac{1}{e^{\beta(\omega_k\mp \mu)}-1}.
\end{equation}
Keeping only the medium contribution, and defining $\Delta \Xi_{m}\equiv  \Xi_{m} - \Xi_0$, 
\begin{equation}
\label{deltam}
\Delta  \Xi_{m}/\Xi_0= \frac{e^2V}{2} \int\frac{d^3q}{(2\pi)^3} \sum_{n_q=-\infty}^{\infty} \frac{{\cal J}_{m} (q_\mu)}{q^2},
\end{equation}
with ${\cal J}_{m}$ conveniently written as
\begin{equation}
\label{calJ}
{\cal J}_{m}={\rm Re}\!\!\int\!\! \frac{d^3p}{(2\pi)^3}\frac{\bar{n}(p)}{\omega_p}\left[2 - \frac{q^2}{q^2+2pq} - \frac{4m^2}{q^2+2pq}\right].
\end{equation}
As before, we subtract the $T=0$ part of Eq.(\ref{deltam}), and call the result $\Delta \Xi^*/\Xi_0$. We will now compute expressions for pressure, density, and condensation temperature in the UR and NR limits, and present numerical results for those quantities in the relativistic case.

In the UR limit, $T>>m$ and/or $\eta >> m^3$, the last term in (\ref{calJ}) does not contribute to the grand partition function in leading order (we have checked this numerically). We perform the sum over $n_q$ in (\ref{deltam}), and the integral over angles in (\ref{calJ}), analytically. The real part of the integral over the radial momentum $p$, and the final radial integral over $q$ may also be computed analytically, yielding
\begin{equation}
\Delta  \Xi^*/\Xi_0= \frac{e^2 }{60} VT^3.
\end{equation}
The thermodynamic potential is given by
\be
\label{thermopot}
\Omega^*= - {T}\ln \Xi^*=-{T} \ln \Xi_0 - {T} \ln [1+ (\Delta \Xi^*/\Xi_0)].
\ee
As $\Delta \Xi^*/\Xi_0$ is of order $\alpha=e^2/4\pi$, we expand the log
\be
\Delta \Omega^* =  \Omega^* - {\Omega_0} \cong - {T} (\Delta \Xi^* / \Xi_0)= - \frac{\pi \alpha}{15} V T^4.
\ee
In the UR limit, $\Omega_0/V T^4 = -\Gamma (4) \zeta (4)/(3\pi^2) = - \pi^2/45$, so the pressure is 
\be
P^* = P_0 + \Delta P^* = \frac{\pi^2}{45} T^4 + \frac{\pi\alpha }{15} T^4.
\ee

\begin{figure}[ht!]
	\begin{center}
		\includegraphics[width=80mm]{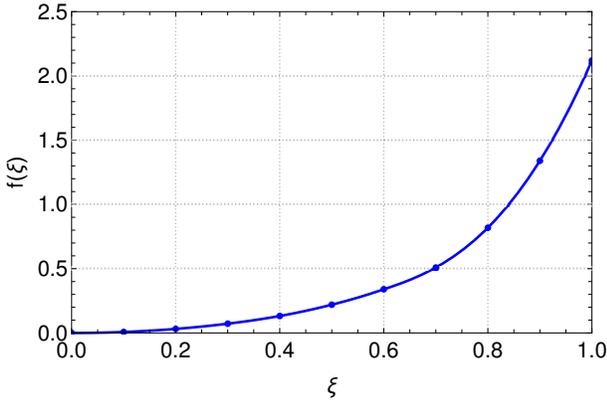}
\caption{\footnotesize Numerical calculation of $f(\xi)$, a function of $\xi=\mu/m$.}
		\label{fig:func_f}
	\end{center}
\end{figure}

We go back to Eq. (\ref{thermopot}) to derive the density
\be
\eta^* = \left ( \frac{\partial P^*}{\partial \mu} \right )_{T,V} = \eta_0 + \Delta \eta^* = \eta_0 - \frac{1}{V} \frac{\partial \Delta \Omega^*}{\partial \mu},
\ee
where $\eta_0$ is the density of the relativistic ideal gas 
\begin{equation}
    \eta_0= \left (\frac{\partial P_0}{\partial \mu} \right )_{T,V} = \!\!\int\!\!\frac{d^3p}{(2\pi)^3}\left[n^+ (p)-n^-(p)\right]. 
\end{equation}
We then use the {\it ansatz}
\be
\label{ansatz}
\eta^* = \eta_0 + \Delta \eta^* = \frac{1}{3} \mu T^2 \left [ 1 + f\left (\xi\right ) \alpha \tau \right].
\ee 
We define $\tau \equiv T/m$, $\xi \equiv \mu/m$, and the function $f(\xi)$, shown in Fig. (\ref{fig:func_f}), which is determined numerically. We take its value for $\mu=m$ to be $f(1)= \gamma$. Numerically, $\gamma = 2.14 \cong 2 \pi/3 $.

In order to obtain $T_c$, the temperature where Bose-Einstein condensation sets in, we use $\mu = m$ and $\eta = \eta^*$
\be
\label{critdens}
\eta = \eta^* = \frac{1}{3}m^3 \tau_c^2 \left [ 1 + \gamma \alpha \tau_c  \right],
\ee
which may be written  
\be
\tau_c^2 \left [ 1 + \gamma \alpha \tau_c \right] = 3\eta/m^3 \equiv N_{\rm u}.
\ee
Setting $\tau_c = {N_{\rm u}}^{1/2} - \Delta \tau$, to first order in $\alpha$
\be
\tau_c =  {N_{\rm u}}^{1/2} - \frac{\gamma}{2} \alpha N_{\rm u},
\ee 
where ${N_{\rm u}}^{1/2}$ is the ideal gas value, and $\alpha N_{\rm u} = (e/4\pi) Q$, with $Q \equiv e \Delta N$ being the total EM charge.

In the NR limit, $T << m$ and/or $\eta << m^3$, we again perform the sum over $n_q$ and integrate over angles. The leading term is now
\be
\Delta \Xi^*/\Xi_0 = \frac{\alpha\tau}{3\pi}  V \eta_0,
\ee
where $\eta_0$ is the nonrelativistic expression for the particle density of the ideal gas (antiparticles are suppressed by $e^{-m/T}$)
\be
\label{etaNR}
\eta_0 (\nu) = \frac{(mT)^{3/2}}{\sqrt{2}\,\pi^2}\int_0^\infty dz \frac{z^{1/2}}{e^{z - \nu} - 1},
\ee
with $\nu \equiv (\mu - m)/T < 0$.
 The shift in the thermodynamic potential with respect to the ideal gas value is
\be
\Delta \Omega^* = - T (\Delta\Xi^*/\Xi_0) = - \frac{\alpha\tau}{3\pi}T V \eta_0.
\ee
However, since $\partial \Omega_0/\partial \nu = - TV \eta_0 $
\bea
&& \Omega^*(\nu) = \Omega_0 (\nu)  + \frac{\alpha\tau}{3\pi} \frac{\partial \Omega_0}{\partial \nu}(\nu) = \Omega_0 (\nu_{\rm em}) ,\\
&& \nu_{\rm em}(\nu, \tau) = \nu + \frac{\alpha \tau}{3\pi}.
\eea
Therefore, the effect of the electromagnetic quantum fluctuations is to increase the chemical potential  by an amount proportional to $(\alpha \tau) T$. They also lead to an increase in the pressure
\be
\Delta P^* = P_0 (\nu_{\rm em}) - P_0 (\nu) = \frac{\alpha \tau}{3\pi} T \eta_0.
\ee

The nonrelativistic density is 
\be
\eta^* (\nu, \tau)=  \frac{1}{T}\frac{\partial P^*}{\partial \nu} =  \frac{1}{T}\frac{\partial P_0}{\partial \nu_{\rm em}} =  \eta_0 (\nu_{\rm em}, \tau).
\ee

Condensation will take place whenever $\nu_{\rm em}=0$, which means $\nu_c = -(\alpha\tau_c)/3\pi$. Then, the total density is given by $\eta =  \eta^* (\nu_c, \tau_c) = \eta_0 (0, \tau_c)$. Since $\nu_c << 1$, we follow \cite{Landau-Lifchitz} to compare the condensation temperature $T_c$ with the temperature $T_0$ of an ideal gas with density $\eta$ and $\nu=\nu_c$
\be
\label{TcNR}
\eta = \zeta(3/2)\left (\frac{m T_0}{2\pi}\right )^{3/2} \left [1 - \frac{\sqrt{4\pi |\nu_c|}}{\zeta(3/2)}\right ].
\ee
Since $\eta = \zeta(3/2) [m T_c/2\pi]^{3/2}$, Eq. (\ref{TcNR}) leads to
\be
\tau_c = \tau_0 \left [ 1 - \frac{2}{3}\frac{\sqrt{4\pi |\nu_c|}}{\zeta(3/2)}\right ].
\ee
Introducing $N_{\rm n} = [(2\pi)^{3/2}/\zeta(3/2)] (\eta/m^3)$,
\bea
&& \tau_c = {N_{\rm n}}^{2/3} - C \alpha^{1/2} {N_{\rm n}},\nonumber \\
&& C = \frac{4}{3\sqrt{3} \zeta(3/2)}= 0.295.
\eea
Again, the shift is proportional to the EM charge $\alpha^{1/2} {N_{\rm n}} = (1/2\sqrt{\pi}) Q$.

The critical temperature is always lower than the ideal gas value, an indication that electromagnetic repulsion acts against condensation. In the ultra-relativistic and nonrelativistic cases, this shift is related to an electromagnetic increase in the chemical potential proportional to $\alpha$
\be
\mu_{\rm em} = \mu + g(\xi, \tau) \alpha T.
\ee
We have $g(\xi, \tau) = \tau/3\pi$ for $\tau << 1$ (NR), and $g(\xi, \tau)= \xi f(\xi)$ for $\tau >> 1$ (UR), with $f(\xi)$ defined in Eq.(\ref{ansatz}). Electromagnetic repulsion is also responsible for an increase in pressure when compared to that of the ideal gas. In the UR and NR limits,
\bea
&& {\Delta P^*_{\rm u}}/{m^4} = \frac{\pi}{15} \alpha \tau^4, \\
&& {\Delta P^*_{\rm n}}/{m^4} = \frac{1}{3\pi} \alpha \tau^2 \frac{\eta_0}{m^3} = \frac{\zeta(3/2)}{6 \pi^{2} \sqrt{2\pi} } \alpha \tau^{7/2}.
\eea
As for the increase in density, in the UR limit, we have
\be
\Delta \eta^*/m^3 =\frac{\xi f(\xi)}{3} \alpha  \tau^3.
\ee
In the NR limit, $\Delta \eta^* = \eta^* (\nu, \tau) - \eta_0 (\nu, \tau) =  \eta_0 (\nu_{\rm em}, \tau) - \eta_0 (\nu, \tau)$. Using Eq.(\ref{TcNR}), we obtain
\be
\Delta \eta^*/m^3 = \frac{1}{6 \pi^2  \sqrt{2|1 -\xi|} } \alpha \tau^{3}.
\ee
\begin{figure}[ht!]
	\begin{center}
		\includegraphics[width=80mm]{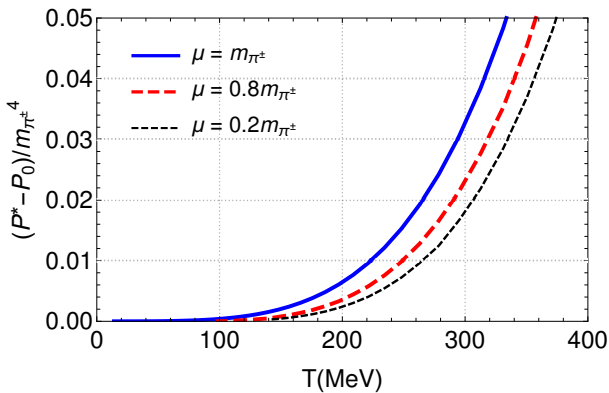}
\caption{\footnotesize Change in the pressure as a function of temperature. The charged pion mass $m_{\pi^\pm}=139.6 {\rm MeV}$ was used for $m$.}
		\label{fig:pressure}
	\end{center}
\end{figure}

\begin{figure}[ht!]
	\begin{center}
		\includegraphics[width=80mm]{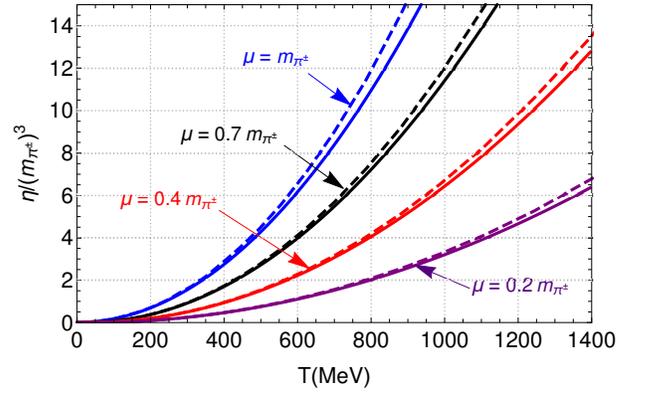}
\caption{\footnotesize Particle density of the ideal gas (full line) and of the gas with electromagnetic interaction (EM gas - dashed line) as a function of temperature for some values of the chemical potential.}
		\label{fig:density}
	\end{center}
\end{figure}
\begin{figure}[ht!]
	\begin{center}
		\includegraphics[width=80mm]{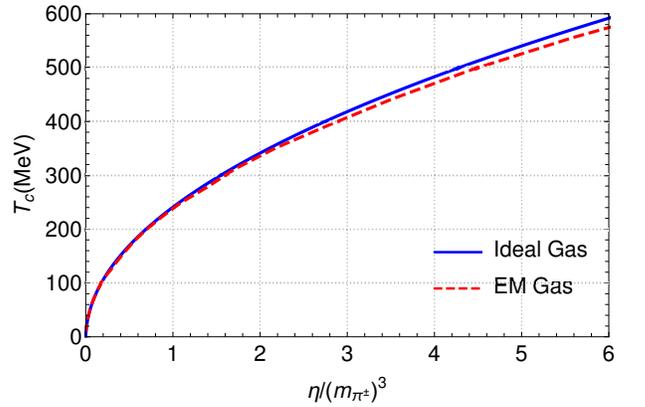}
\caption{\footnotesize Condensation temperature as a function of particle density for the ideal gas and for the gas with electromagnetic interaction.}
		\label{fig:tc}
	\end{center}
\end{figure}

For the relativistic case, the shift in pressure and density as functions of temperature for some values of the chemical potential are shown in Figs.(\ref{fig:pressure}) and (\ref{fig:density}), respectively. Fig.(\ref{fig:tc}) shows the critical temperature as a function of the density. In all the figures, we have taken $m$ to be the charged pion mass $m_{\pi^\pm} =139.6 {\rm MeV}/c^2$.

It is clear from the numerical calculations that the pressure and the density increase with respect to the ideal gas values, whereas the condensation temperature decreases, all this as a consequence of the electromagnetic repulsion that sets in via quantum fluctuations. The increase in pressure and density is expected because the gas will experience a repulsive interaction, since we are fixing the electromagnetic (EM) charge $Q = e\Delta N$, proportional to the
number of particles minus the number of antiparticles. For positive chemical potential $\mu>0$, there are more
particles than antiparticles. At low temperatures, no antiparticles
will be produced from the vacuum, so that we are left with a system of particles (positive net particle number), all with the same EM charge $e$, that repel each other. As we increase the temperature above $2mc^2$,
zero-charge particle-antiparticle pairs will be continuously created from and annihilated into the vacuum, but this does not change the net charge, and the system still consists of same sign charges that experience
a net repulsion. 
Although the results are order $\alpha$ corrections, as we move into the UR regime, those corrections become more relevant. However, we cannot trust our first order calculation for values of $\tau=T/m$ close to, or larger than, $1/\alpha$. 

Since our calculations show that the first order $\mathcal{O}(e^2)$ of the expansion gives a small, yet relevant, contribution in the ultra-relativistic limit, higher order $\mathcal{O}(e^4)$ corrections will lead to even smaller contributions that can be neglected in the calculation of physical quantities such as the pressure, density, and critical temperature of condensation.

We point out that temperatures in the graphs do not exceed $\tau \sim 10< 137$, spanning a region where our approximations should hold. Note that the increase in pressure already reaches $5\%$ for values of $\tau \sim 2-3$. As for the increase in density, it may reach $9\%$ for $\tau \sim 6$, whereas the decrease in condensation temperature reaches $5\%$ for $\tau \sim 6$. In order to detect changes of the order of $5-10 \%$, one would have to search for physical scenarios with temperatures corresponding to $\tau= 2 - 10$, so that relativistic effects become appreciable still within the validity of our computations.

In the inner core of a neutron star, where pion condensation may occur \cite{loewe, jens, fraga}, densities exceed $10^{14}\,{\rm g/cm}^3$, with number density $\rho > 0.4 \, {\rm fm}^{-3}$, and pressure $P > 10^{33}\,{\rm dyn/cm^2}$. In our numerical calculations, the shift in condensation temperature shown in Fig.(\ref{fig:tc}) becomes appreciable for number densities $\eta/{(m_{\pi^{\pm}})^3} > 2$, or $\eta > 1.4 \, {\rm fm}^{-3}$, and temperatures where $k T_c > 350\, {\rm MeV}$ or $T_c >  4 \times
10^{12}\, \rm{K}$. Such temperatures $T\sim 10^{12}\, {\rm K}$ are found in the center of neutron stars.

In conclusion, we have shown that EM quantum fluctuations increase the pressure and the density of the gas with respect to ideal gas values by amounts proportional to $\alpha$ times a power of $\tau$. The condensation temperature is, however, lowered by a correction proportional to the charge $Q = e \Delta N$. The shifts become more relevant (of the order of a few percent) the more relativistic is the system. The combination of quantum and relativistic effects in dense hot gases leading to deviations from ideal gas behavior should also be present whenever other conserved charges interact via the exchange of their vector boson carriers, as in Quantum Flavordynamics or Quantum Chromodynamics, for example. We plan to investigate this in the near future.

\acknowledgments

The authors would like to thank the Brazilian Agencies CNPq and CAPES for partial financial support.


\begin{thebibliography}{99}
\bibitem{measure} $d\Sigma[A] \equiv [dA_\nu]\, {\rm det}[\delta {\mathcal G} /\delta {\Lambda}]\, \delta [{\mathcal G}]$. The determinant is the Jacobian of the gauge transformation $A_\nu \rightarrow A_\nu - \partial_\nu \Lambda$, and the functional delta implements a gauge condition, such as ${\mathcal G} [A]=\partial_\nu A_\nu = 0$.
\bibitem{finite size} The finite $T$ free photon propagator, for $r=|\vec{x} - \vec{y}|$, $r_4= |x_4 - y_4|$, is
\be
G(r, r_4) = \frac{T}{8\pi r}\frac{\sinh (2\pi T r)}{\sin ^2 (\pi T r_4) + \sinh ^2 (\pi T r)}. \nonumber
\ee
Used in Eq.(\ref{grand}) for the disconnected term, it gives $\sim Q^2/R T$, $V=\frac{4\pi}{3} R^3$, $Q= e \Delta N$. Note that, at $T=0$, the expression for the propagator properly reduces to $G(x-y) = 1/4 \pi^2 (x - y) ^2$.
\end{thebibliography}
\end{document}